# Thermal Neutron Measurements with an Unpowered, Miniature, Solid-State Device

**Tim Hossain and Clayton Fullwood**

Cerium Laboratories, 5204 E Ben White Blvd, Austin, TX 78741

**Will Flanagan[1,*], Peter Hedlesky, John Rabaey, Steven Block[2]**

The University of Dallas, Department of Physics, 1845 E Northgate Dr, Irving, TX 75062

**Aidan Medcalf**

HighPoint Design, 2030 Century Center Blvd, Irving, TX 75062

**Tracy Tipping**

The University of Texas at Austin, Nuclear Engineering Teaching Laboratory, 11131 Creativity Trail, Austin, TX 78758

[1]Also at Army Applications Lab, 701 Brazos St, Austin, TX 78701

[2]Also at Hiller Measurements, 14155 E Hwy 290, Dripping Springs, TX 78620

[*]Corresponding Author: wflanagan@udallas.edu

## Abstract

A prototype neutron detector has been created through modification to a commercial non-volatile flash memory device. Studies are being performed to modify this prototype into a purpose-built device with greater performance and functionality. This paper describes a demonstration of this technology using a thermal neutron beam produced by a TRIGA research reactor. With a 4x4 array of 16 prototype devices, the full widths of the beam dimensions at half maximum are measured to be 2.2x2.1 $cm^2$.

## 1. Introduction

The field of neutron detection was born in 1932 with the observation of recoiling protons resulting from elastic scattering with the as-of-yet undiscovered neutron [1,2]. Whether these recoiling protons are detected through cloud chamber photography, ionization, scintillation, or any other means, these techniques remain the dominant method of fast neutron detection. At the other end of the energy spectrum, detection of fission fragments from $^3$He, $^6$Li, $^{10}$B and fissile isotopes present the dominant method of measuring thermal (slow) neutrons. In both cases, these detectors typically require power and data acquisition [3,4]. The technology presented here requires neither, allowing novel applications and a genuine leap in detector miniturization.

The prototype presented is modified from a SONOS memory device [5]. In a SONOS (more precisely silicon – silicon dioxide – silicon nitride – silicon dioxide – silicon) device, the silicon nitride layer stores charge to designate a logical 0 or 1. This layer borders an insulating borophosphosilicate glass (BPSG) layer. The enriched boron-10 within the prototype BPSG layer undergoes neutron fission ($n + {}^{10}B \rightarrow {}^{4}He + {}^{7}Li$) with the highly-ionizing fission isotopes neutralizing the stored charge, decreasing the capacitor voltage. The non-volatile nature of SONOS devices allows this process to occur and be preserved even when the prototype is unpowered. The silicon dies are packaged within a 11x13$mm^2$ ball grid array (BGA) footprint and contain 2,147,483,648 ($2^{31}$) charge storage elements. The critical dimension of this prototype is 65nm.

The sensitivity of memory devices to neutron radiation as a result of boron-10 content has been studied since 1995 [6,7]. Utilizing this fact to create a novel neutron detector was conceived shortly thereafter [8]. Similar observation of soft errors and intentional creation of a neutron detector have been studied in additional families of memory devices such as the volatile SRAM architecture [9,10].

## 2. Experimental Setup

A 4x4 array of 16 prototype devices was placed along a thermal beam at The University of Texas at Austin Nuclear Engineering Teaching Laboratory TRIGA research reactor [11]. In order to create a simple detector footprint, the prototypes were left unconnected and unpowered, suspended with a 3D printed fixture. Prior to exposure, all bits were set to a logical 0, corresponding to roughly 5.7V. To confirm the beam flux and shape with an independent source, a 4x4 array of gold foils with the same footprint was placed in an identical fixture [Figure 1]. The gold foil fixture was mounted in the beamline after the prototype fixture, maintaining the same exposure conditions.

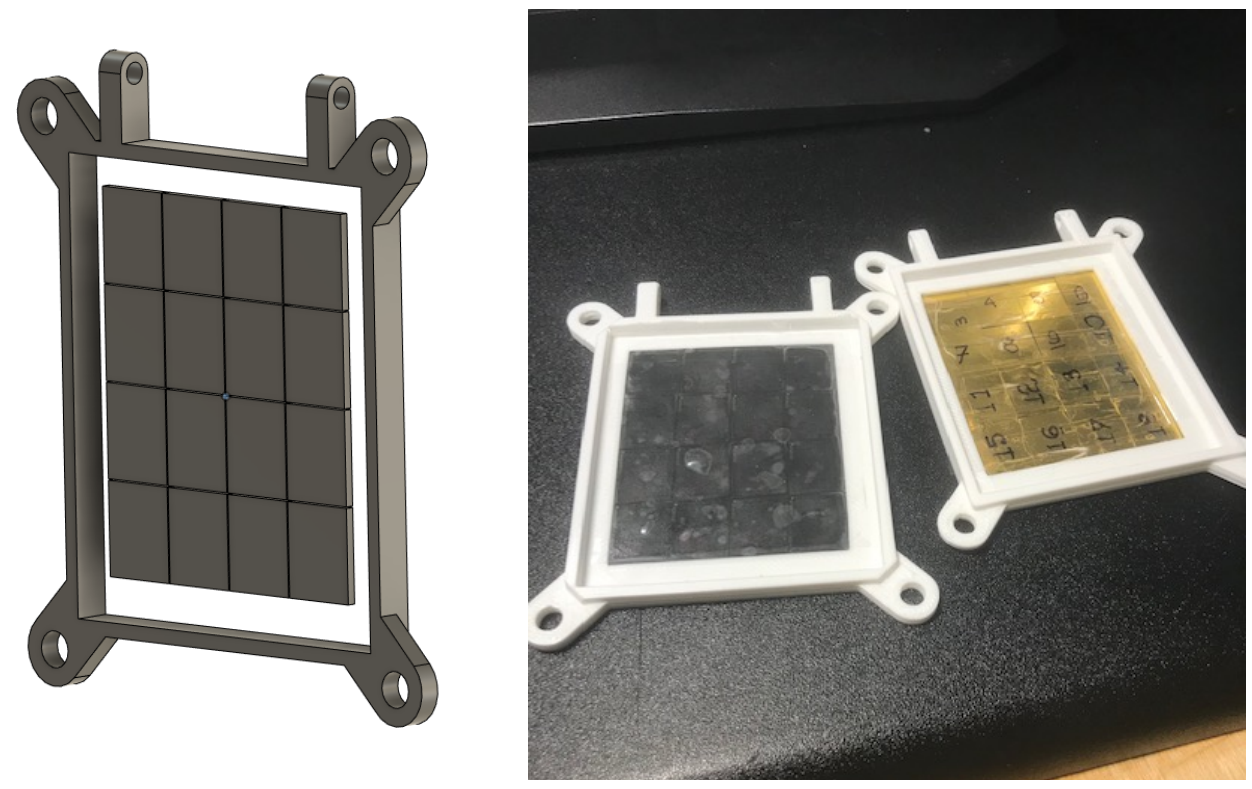

**Figure 1** Left: A 3D model of the experimental fixture shown with a 4x4 array of BGA-packaged prototype detectors. Right: A photograph of the empty prototype fixture and gold foil fixture, prior to exposure.

## 3. Results

The voltage of each bit was measured to within 100mV before and after the exposure. An exponential tail of charge loss is clearly visible for all 16 prototypes [Figure 2]. The amount of charge loss in each individual charge storage element is given in Figure 3. The number of bits with a voltage shift at or beyond 300mV is plotted on its physical position for each sector of 1,048,576 ($2^{20}$) bits. The full widths of the beam profile dimensions at half maximum are measured to be 2.2cm in the horizontal (x) dimension and 2.1cm in the vertical (y) dimension. Further granularity of the plot is possible by using the physical position of each bit, though it does not increase the precision of this particular measurement.

The neutron flux was computed at each gold foil post-exposure using neutron activation analysis. The compatible beam profile shapes are shown in Figure 4.

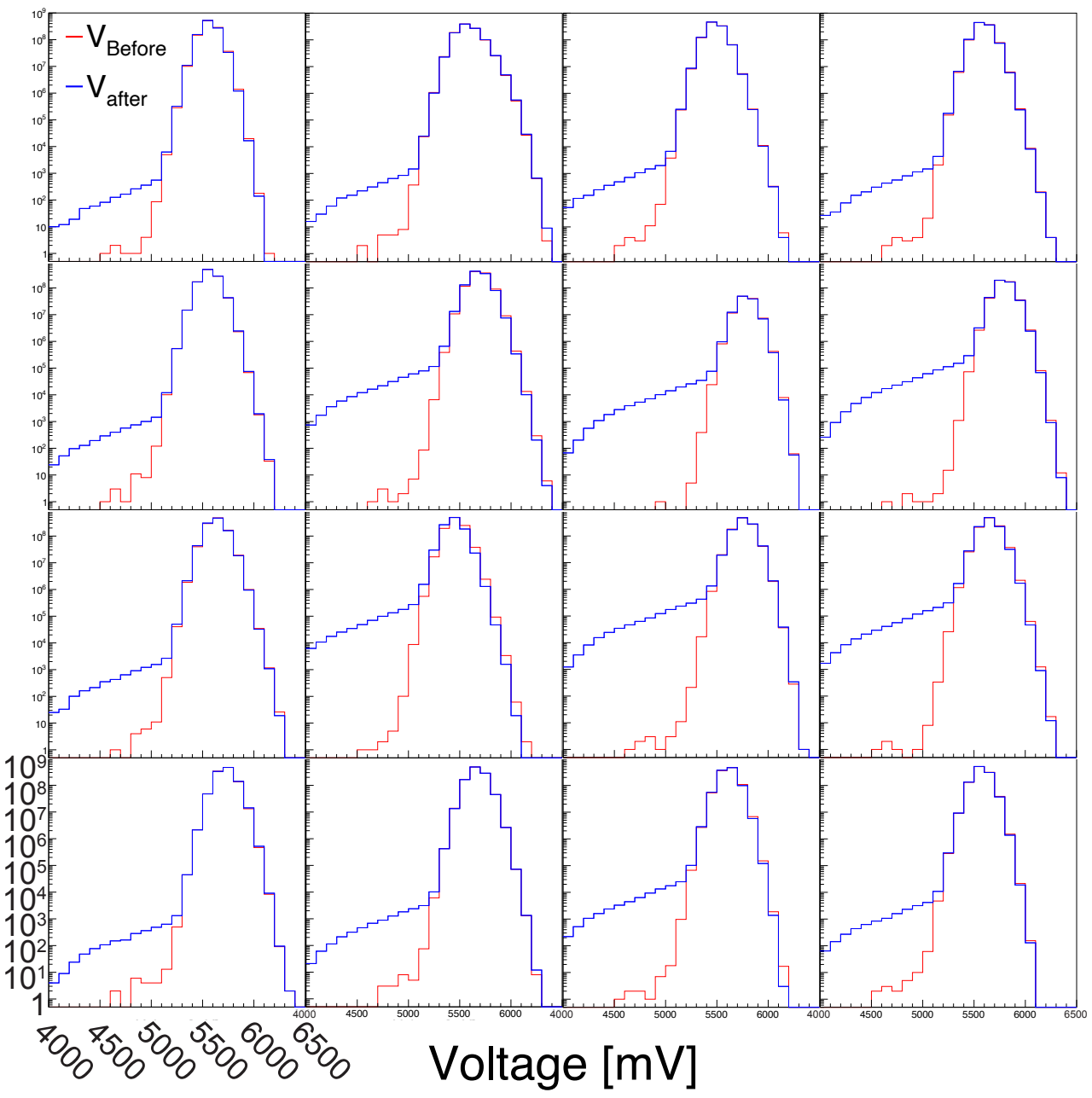

**Figure 2** The distributions of individual bit voltages before (red) and after (blue) exposure to the thermal neutron beam for all 16 prototype detectors.

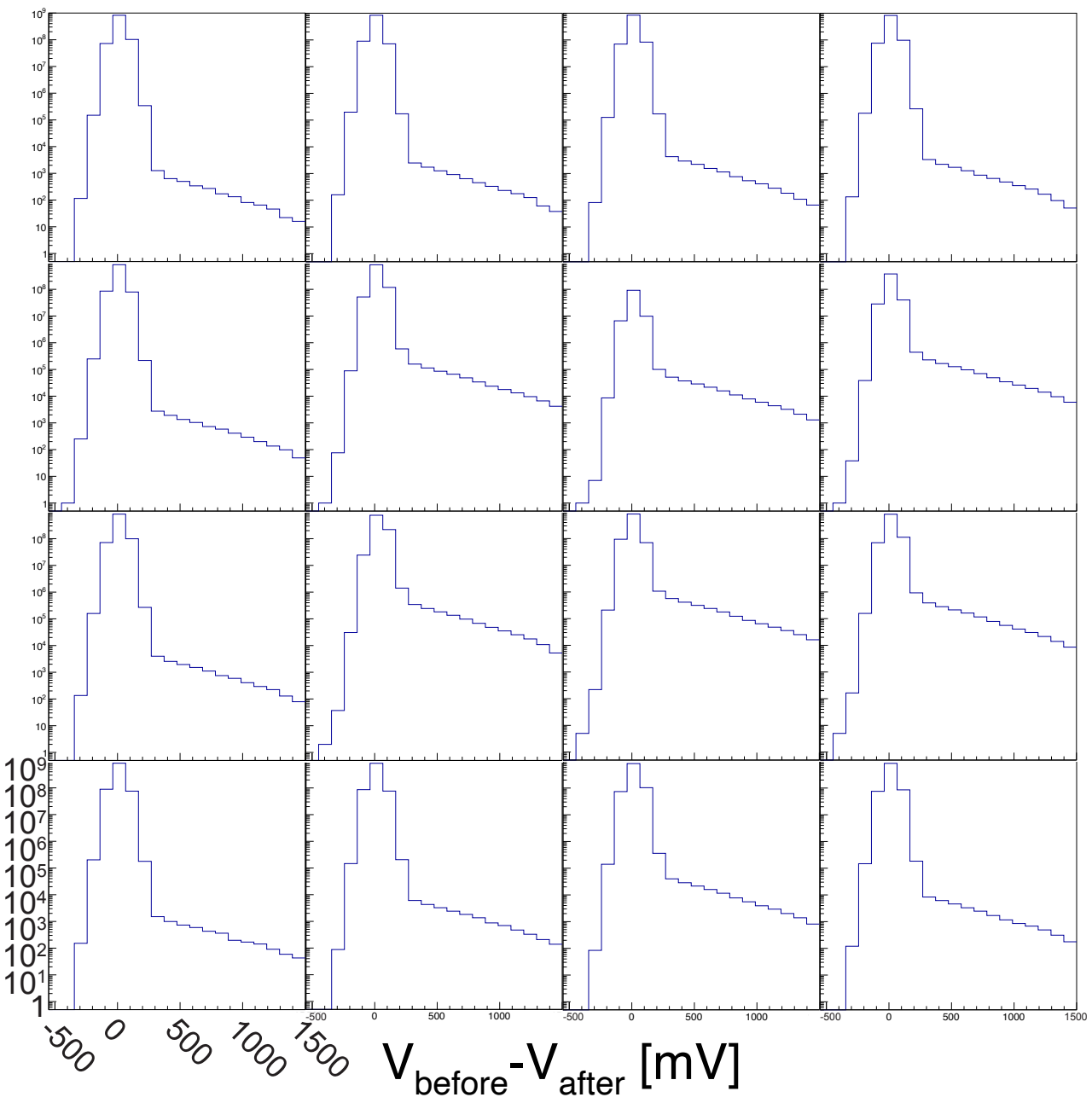

**Figure 3** The distribution of post-exposure voltage subtracted from pre-exposure voltage for each individual bit for all 16 prototype detectors.

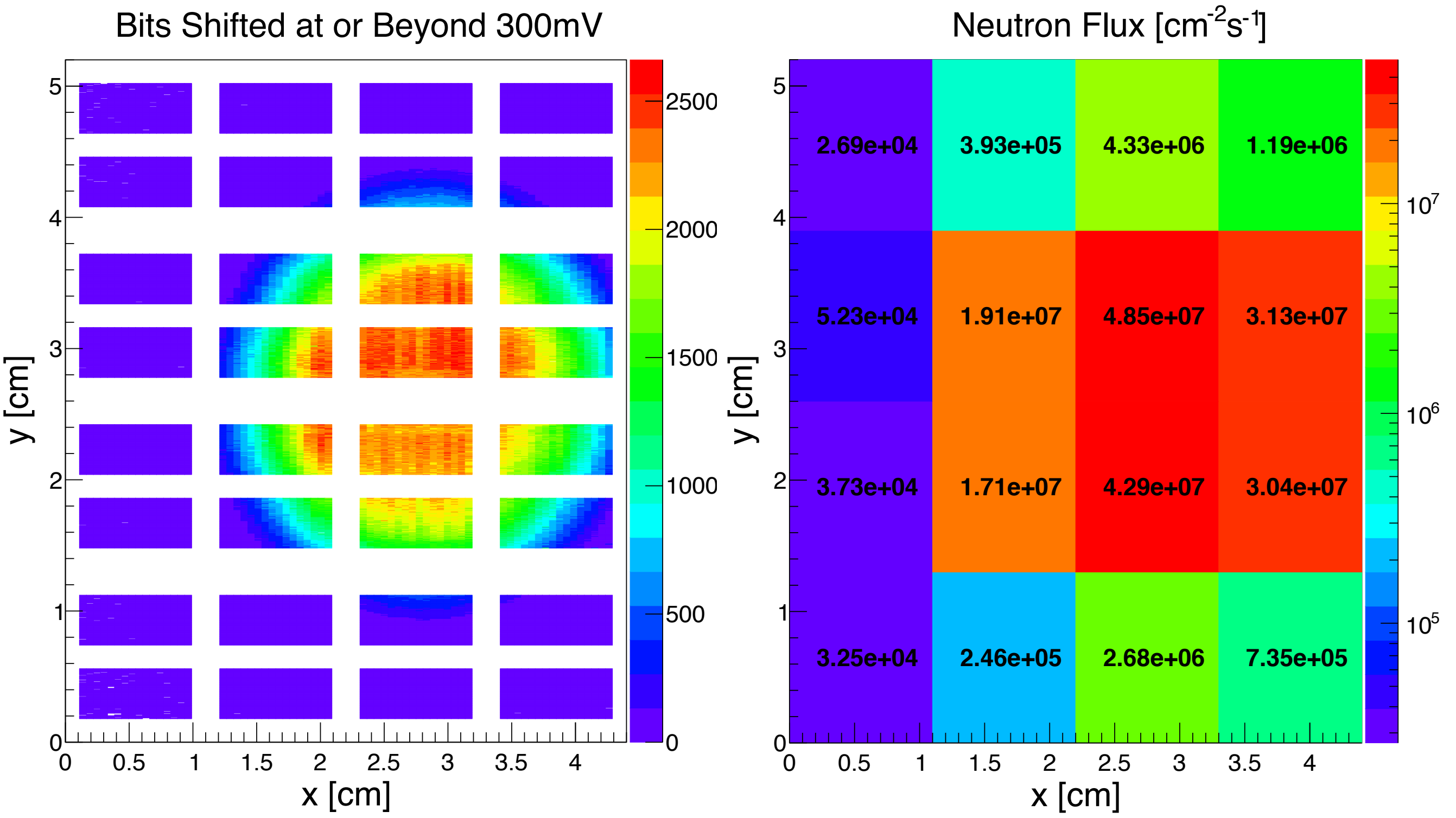

**Figure 4** Left: The number of bits with a voltage shift beyond 300mV is plotted over physical location within the fixture. Each prototype detector has two blocks of charge storage elements, creating the structure shown. Each pixel within the plot represents $2^{20}$ bits. Right: The flux computed by neutron activation analysis for each of the 16 gold foils.

## 4. Conclusion

This device holds promise in revolutionizing neutron imaging in the sub-micron realm [12, 13, 14, 15]. Though outside of the project's original scope, a more precise position resolution measurement will be undertaken. The team is preparing for an integrated circuit redesign to encompass various performance improvements inherent to a purpose-built device. In parallel, applications are being demonstrated of these prototype detectors as (passive, stand-alone) neutron-sensitive dosimeters and (active, powered) real-time neutron counting.

## Acknowledgements

This work was supported by a Phase II contract from the Air Force Small Business Technology Transfer (STTR) program (contract number FA864919PA076) and the University of Dallas Donald A. Cowan Physics Institute. The authors also acknowledge the Texas Advanced Computing Center (TACC) at The University of Texas at Austin for providing grid computing resources that have contributed to the research results reported within this paper. Finally, the views presented are those of the authors and do not necessarily represent the views of DoD or its components.